\pdfoutput=1

\documentclass[aps,prl,2pt,showpacs,superscriptaddress]{revtex4}
\usepackage{amsmath}
\usepackage{latexsym}
\usepackage{amssymb}
\usepackage{graphics,epstopdf}
\usepackage{slashed}
\usepackage[colorlinks=true, citecolor=blue, urlcolor=blue ]{hyperref}
\begin{document}
\title{Quantum Restoration of Broken Symmetry in one Dimensional Loop Space}

\author{Pinaki Patra}
\email{monk.ju@gmail.com}
\affiliation{Department of Physics,University of Kalyani,India-741235}

\author{Tanmay Mandal}
\email{mandal.tanmay3@gmail.com}
\affiliation{Department of Physics,University of Kalyani,India-741235}

\author{Jyoti Prasad Saha}
\email{jyotiprasadsaha@gmail.com}
\affiliation{Department of Physics,University of Kalyani,India-741235}

\begin{abstract}
For 1 Dimensional loop space, a nonlinear nonlocal transformation of fields is given to make the action of the self-interacting quantum field to the free one. A specific type of Classically broken symmetry is restored in Quantum theory. 1-D Sine Gordon system and Sech interactions are treated as the explicit example.
\end{abstract}

\maketitle
\section{Introduction}
Though Quantum theory and Classical theory are completely different from the philosophical point of view, it is a long time belief among many people(including the present authors) that the Classical property of a system and Classical limit of the corresponding Quantum system will be identical\cite{Bohr}. In Quantum field theory, one has to estimate the probability of occurrence of some particular event. So, it is necessary to have a knowledge about the probability measure of the concerned scenario. Sometimes, for the description of a system, the knowledge of the Probability measure is sufficient. But, for the interacting theory, it is difficult(most of the cases impossible) to have the exact knowledge about the system. Recently, Belokurov et. all \cite{Namsrai,Belokurov,Shavgulidze} gave some idea to handle the interacting system for 1-Dimensional case. With the help of a nonlinear nonlocal transformation of the fields, they explicitly studied the $\phi^4$ interaction theory and were able to transform the theory as a free field theory. But, they were landed in a strange situation. The Classical theory and the Classical limit of the corresponding Quantum theory are completely different before and after the transformation. As a result of the transformation singularity property of two fields (the original and and the transformed one) are completely different. In this article, we have proposed a nonlinear nonlocal transformation of the fields for the general functional form of interaction term and as a result we got similar stand point as of Belokurov et. all.  Also, it can be observed that sometimes the original classical system is symmetric with respect to the transformation $T(\phi)=-\phi$; while the symmetry is not preserved for the transformed classical system. But, as for the Quantum system, we have to deal only with the functional Weiner measure; from the discussion of this article we shall see that the functional measure is invariant with respect to the transformation. In this sense the classically broken symmetry is restored in Quantum case. \\
In the next section, we have described the general formalism and then subsequently two special cases are studied.
\section{General Construction}
If $\mathcal{B}$ is the Borel $\sigma$-field on $\Omega=\mathcal{C}[0,1]$ and $\mathcal{B}_t$ be the $\sigma$-field generated by $\phi(s)$ for $0\leq s\leq t$ and if we consider the Wiener measure $P$ on $(\Omega,\mathcal{B})$, then one can identify that $\phi(t)$ is a martingle with respect to $(\Omega,\mathcal{B}_t,P)$. So, for a meromorphic function $f(\phi(t))$, one can consider the Wiener measure \cite{Srinivasa}
\begin{equation}
P=\exp\{\int_{t=0}^{t=1}[-\frac{1}{2}\dot{\phi}^2(t)-\frac{\lambda^2}{4}f^2(\phi(t))]dt\}d\phi
\end{equation}
where $\lambda$ is the parameter, the dimension of which depends on the form of $f(\phi)$. Now, for our concerned loop space $\Omega$, we can write the following proposition.\\
{\bf Proposition 1:}
For the  nonlocal transformation,
\begin{equation}
\chi(t)=\phi(t)+\frac{\lambda}{\sqrt{2}}\int_{\tau=0}^{\tau=t}f(\phi)d\tau
\end{equation}
the following equality holds
\begin{eqnarray}
\int \exp (-\frac{1}{2}\int_{t=0}^{t=1}\dot{\chi}^2dt) d\chi= \int \exp \{\int_{t=0}^{t=1}[-\frac{1}{2} \dot{\phi}^2(t) -\frac{\lambda^2}{4}f^2(\phi(t))  +\frac{\lambda}{2\sqrt{2}}\frac{\partial^2h}{\partial\phi^2}] dt - \frac{\lambda}{\sqrt{2}}(h(\phi(1))-h(\phi(0)))\}d\phi \\ \mbox{if},h(\phi)=\int f(\phi)d\phi. 
\end{eqnarray}
{\bf Proof:}
To prove the proposition it is sufficient to note that with the help of the well known procedure of the Ito-Stochastic integration \cite{Srinivasa}, we can write
\begin{equation}
\int_0^1 \dot{\phi}f(\phi)dt=h(\phi(1))-h(\phi(0))-\frac{1}{2}\int_0^1\frac{\partial^2h}{\partial\phi^2} dt\;;h(\phi)=\int f(\phi)d\phi.
\end{equation}
After noting the above-mentioned result, the proof is rather straightforward. We have to just use the above-mentioned transformation of fields and write $\dot{\chi}=\dot{\phi}+\frac{\lambda}{\sqrt{2}}f(\phi)$ and put the expression in the Weiner measure $P$.\\
We should note that, though the above-mentioned equality holds, $\phi(t)$ and $\chi(t)$ belongs to  different functional space and the normalization of the spaces of $\phi$ and $\chi$ are different. That is if $\chi(t)\in \mathcal{C}[0,1]$ and $\phi(t)\in X$, then $X\neq \mathcal{C}[0,1]$; in fact, the space of $\phi$ is more singular than that of $\chi$. In \cite{Belokurov}, Belokurov et.all have constructed the detail singularity property for $f(\phi)=\phi^4$. From that discussion we can intuitively argue that, the space of $\chi$ is more singular than that of $\phi$. But, the exact study of singularity for the general case $f(\phi)$ is beyond the knowledge and capability of the present authors.\\
If we set $\exp\{-\frac{1}{2}\dot{\chi}^2\}d\chi$ as the measure on $\mathcal{C}[0,1]$; then obviously, $\exp\{-\frac{1}{2}\dot{\phi}^2\}d\phi$ is not the measure in $X$. But, we can set, $\exp \{\int_{t=0}^{t=1}[-\frac{1}{2} \dot{\phi}^2(t) -\frac{\lambda^2}{4}f^2(\phi(t))+\frac{\lambda}{2\sqrt{2}} \frac{\partial^2h}{\partial\phi^2}] dt - \frac{\lambda}{\sqrt{2}}(h(\phi(1))-h(\phi(0)))\}d\phi$ as the measure on $X$.\\
The above-mentioned identification leads to the following  interesting proposition-
{\bf Proposition 2:} If the function $f(\phi)$ is even function of $\phi$, then the Classical action and the classical limit of the corresponding Quantum action shows different symmetry property under the symmetry transformation $T(\phi)=-\phi$. Moreover, if $f(\phi)$ is odd function of $\phi$, then the Classical action and the classical limit of the corresponding Quantum action shows similar symmetry property under the symmetry transformation $T(\phi)=-\phi$.\\
{\bf Proof:} For the classical action
\begin{equation}
A_{old}=\int_{t=0}^{t=1}[\frac{1}{2}\dot{\phi}^2(t)+\frac{\lambda^2}{4}f^2(\phi(t))]dt
\end{equation}
the equation of motion takes the form
\begin{equation}
\ddot{\phi}=\frac{\lambda^2}{2}f(\phi)\frac{\partial f(\phi)}{\partial\phi}
\end{equation}
which is symmetric under the transformation $T(\phi)=-\phi$ for any of the two cases of odd and even functional form of $f(\phi)$. \\
But, for the classical action $A_{new}=A_{old}- A_{extra}$; where,
\begin{equation}
A_{extra}= \frac{\lambda}{2\sqrt{2}}\int_{t=0}^{t=1} \frac{\partial f(\phi)}{\partial\phi}dt
\end{equation}
the equation of motion
\begin{equation}
\ddot{\phi}=\frac{\lambda^2}{2}f(\phi)\frac{\partial f(\phi)}{\partial\phi}-\frac{\lambda}{2\sqrt{2}} \frac{\partial^2f(\phi)}{\partial\phi^2}
\end{equation}
is not symmetric for the case of even function $f(\phi)$ under the transformation $T$ due to the presence of second derivative term (As, the second derivative of the even function remains even function). Whereas, for the case of odd function $f(\phi)$, the symmetry is not broken(As, the second derivative of the odd function remains odd function).\\
Now,the corresponding Quantum theory deals with the functional measure $\int \exp\{-A_{new}(\phi)\}d\phi$. Because of the equality of the functional measure as stated in 'Proposition 1'; the symmetry is restored in Quantum case even after the nonlinear nonlocal transformation (2). \\
The above-mentioned argument completes the proof. If we can see this by the following way we shall get another interesting property.\\
The equation of motion for the action of the left side of the equation (3), i.e $A_\chi=-\frac{1}{2}\int \dot{\chi}^2dt$ is
\begin{equation}
\ddot{\chi}=0\Longrightarrow \dot{\chi}=\mbox{constant}=\alpha (\mbox{let})
\end{equation}
which implies (by virtue of the transformation (2))
\begin{equation}
\dot{\phi}=\alpha-\frac{\lambda}{\sqrt{2}}f(\phi)
\end{equation}
Also, from the equation of motion in (6) we can get
\begin{equation}
\dot{\phi}=\pm \beta f(\phi)\;;\mbox{where},\beta=\mbox{some\; constant}
\end{equation}
which can be identified with equation (10),if we set $\alpha=0$ (which we can always choose, as this satisfies the boundary condition) and $\beta=\frac{\lambda}{\sqrt{2}}$. The $\pm$ correspond to two different branches of the solutions. Actually the appearance of the branches are quite natural as we are taking the square root of $f^2(\phi)$ in the equation (11).
The explicit form of the solution is
\begin{equation}
\phi(t)=u^{-1}(-\frac{\lambda}{\sqrt{2}}t+\delta)\;,\delta=\mbox{some\; constant}\;,\int\frac{d\phi}{f(\phi)}=u(\phi)
\end{equation}
But, one can easily identify that equation(8) cannot be reduced in the form as above of equation (11). So, a solution of the form of (12) satisfy (6),but not (8). But, they both corresponds to the Classical limit of the same Quantum system (because, of the equality of the functional measure (3)). So, in the classical case the symmetry is broken; whereas, in Quantum case the symmetry is preserved.\\
In the next section some examples are discussed with the help of this construction.
\section{Examples}
Now we shall discuss the situation for Sine-Gordon system and for Sech interaction. 
\subsection{Sine-Gordon System (odd form of $f(\phi)$)}
As our discussion is confined on a loop space, it is quite natural to think of the analysis for the case of Sine Gordon System\cite{Rajaraman,Ablowitch}.  It is well known that for specific values of the parameter of $Sine$ Gordon system it becomes equivalent to the Thirring model \cite{Coleman}. The one of the most important observation for the Thirring model is though the $Sine$-Gordon equation is the theory of massless scalar field, the $Sine$- Gordon soliton \cite{Coleman} can be identified with the fundamental fermion of the Thirring model.
\\
To obtain the Sine-Gordon equation on $\mathcal{C}[0,1]$, we can consider the action
\begin{equation}
A_{sg}=-\frac{1}{2}\int_{t=0}^{t=1}[\dot{\phi}^2+\frac{\lambda^2}{4}\sin^2\frac{\phi}{2}]dt
\end{equation}
Then the nonlinear nonlocal transformation
\begin{equation}
\chi(t)=\phi(t)+\frac{\lambda}{\sqrt{2}}\int_0^t \sin\frac{\phi(\tau)}{2} d\tau
\end{equation}
leads to the equality of the functional measure
\begin{eqnarray}
\int e^{-\frac{1}{2}\int_0^1\dot{\chi}^2dt} d\chi=\int\exp\{-\frac{1}{2}\int_0^1\dot{\phi}^2 dt- \frac{\lambda^2}{4}\int_0^1\sin^2\frac{\phi}{2}dt  +\frac{\lambda}{4\sqrt{2}}\int_0^1 \cos\frac{\phi}{2}dt +2\frac{\lambda}{\sqrt{2}}(\cos\frac{\phi(1)}{2}-\cos\frac{\phi(0)}{2})\}d\phi
\end{eqnarray}
Though the symmetry property of the old system and the new system(after nonlocal transformation) are the similar with respect to the symmetry transformation $T(\phi)=-\phi$, the two system shows different properties with respect to their solutions. \\
In particular, the equation of motion obtained from the $A_sg$ admits a solution of the form
\begin{equation}
\phi(t)=4\cot^{-1}(ie^{-\frac{it}{2}})
\end{equation}
But, the solution of that form will not be the solution of
\begin{equation}
\ddot{\phi}=\frac{\lambda^2}{8}\sin\phi-\frac{\lambda}{8\sqrt{2}}\sin\frac{\phi}{2}
\end{equation}
which is obtained from the new action after the nonlocal transformation. \\
So, in that sense the symmetry is broken for the Classical case. Whereas, for the Quantum case by virtue of the equality of the functional measure, the symmetry is preserved.
\subsection{Sech Interaction (even form of $f(\phi)$)}
Due to the appearance as a Soliton solution of a large class of integrable nonlinear partial differential equation \cite{Rajaraman,Ablowitch}, Sech interaction is very much interesting among Physicist. One of the interesting feature of this type of interaction is that it may be used as a model of reflection less potential \cite{Kiriushcheva} and also for the case of the interaction of the form of compactly supported function(as an example, the alternate deposition of thin layers of $GaAs$ and $Al_xGa_{1-x}As$\cite{Miller} may be modelled with Sech interaction; as, in that case the lower bound of the interaction is present).\\
Now, for this case we can consider the Action
\begin{equation}
A_{sec}=\int_0^1(\frac{1}{2}\dot{\phi}^2+\frac{\lambda^2}{4}sech^2\phi)dt
\end{equation}
The corresponding Weiner measure is
\begin{equation}
P=\exp\{-\frac{1}{2}\int_0^1\dot{\phi}^2(t)-\frac{\lambda^2}{4}\int_0^1 sech^2\phi(t)dt\}d\phi
\end{equation}
Therefore, according to proposition 1, for the nonlinear nonlocal transformation
\begin{equation}
\chi(t)=\phi(t)+\frac{\lambda}{2}\int_0^t sech\phi(\tau)d\tau
\end{equation}
the following equality holds
\begin{eqnarray}
\int e^{-\frac{1}{2}\int_0^1\dot{\chi}^2dt}d\chi =\int \exp\{\int_0^1[-\frac{1}{2}\dot{\phi}^2- \frac{\lambda^2}{4} sech^2\phi  -\frac{\lambda}{2\sqrt{2}}sech\phi\tanh\phi]dt-\frac{2\lambda} {\sqrt{2}}(\tan^{-1} e^{\phi(1)}-\tan^{-1} e^{\phi(0)})\}d\phi
\end{eqnarray}
Now the equation of motion obtained from $A_{sec}$
\begin{equation}
\ddot{\phi}=-\frac{\lambda^2}{2}sech^2\phi \tanh\phi
\end{equation}
admits a solution of the form (if we set the integration constant with $\frac{\lambda^2}{2}$)
\begin{equation}
\phi(t)=\sinh^{-1}(\frac{\lambda}{\sqrt{2}}t+\beta)\;;\beta=\mbox{constant}
\end{equation}

If we look at the equation of motion after transformation this gives
\begin{equation}
\ddot{\phi}=-\frac{\lambda^2}{2}sech^2\phi\tanh\phi+\frac{\lambda}{2\sqrt{2}}sech\phi(1-2\tanh^2\phi)
\end{equation}
which has not symmetric with respect to the $T(\phi)$ (contrary to the original case).\\
But, as we have already noted in the previous section that Quantum system deals with the functional measure, by virtue of the above-mentioned equality of the functional measure, we can conclude that symmetry is preserved in Quantum case; whereas, Classically symmetry is broken.

\section{Discussion}
Now, after the above construction we can say that the nonlocal transformation of fields may leads to some completely different system. In the article of Belokurov et. all, they point out the well known Haag's theorem which may be helpful to understand this scenario. Even if we completely forget the Haag's theorem, we have to must admit the fact that inclusion of non-locality in the field theory will give some extra interesting information. Nowadays, it is well known that in some cases, the inclusion of non-locality can be modelled as Lorentz invariant CPT violating theory \cite{Chaichian, Masud}. These theory give some additional interesting phenomenon of mass splitting between particle and corresponding antiparticle and may be helpful to justify the recent data analysis results which speculates the mass difference between Neutrino and anti-Neutrino \cite{Experiment1, Experiment2}. Therefore, we can conclude that behind the simple mathematical operations of this article, a deeper Physical Phenomenon may be hiding which may be interesting to be studied in future.
\section*{Acknowledgments :}Pinaki Patra is grateful to CSIR for fellowship support which made this work possible.

\end{document}